\documentstyle[12pt]{article}
\newlength{\extraspace}
\newlength{\extraspaces}
\newcommand{\be}{\begin{equation}
\addtolength{\abovedisplayskip}{\extraspaces}
\addtolength{\belowdisplayskip}{\extraspaces}
\addtolength{\abovedisplayshortskip}{\extraspace}
\addtolength{\belowdisplayshortskip}{\extraspace}}
\newcommand{\ee}{\end{equation}}

\newcommand{\ba}{\begin{eqnarray}
\addtolength{\abovedisplayskip}{\extraspaces}
\addtolength{\belowdisplayskip}{\extraspaces}
\addtolength{\abovedisplayshortskip}{\extraspace}
\addtolength{\belowdisplayshortskip}{\extraspace}}
\newcommand{\ea}{\end{eqnarray}}

\newcommand{\nonu}{\nonumber \\[.5mm]}
\newcommand{\A}{&\!\!\!}

\newcommand{\newsection}[1]{
\vspace{7mm} \pagebreak[3] \addtocounter{section}{1}
\setcounter{subsection}{0} \setcounter{footnote}{0}
\begin{center}
{\large {\bf \thesection. #1}}
\end{center}
\nopagebreak
\medskip
\nopagebreak \hspace{3mm}}

\begin{document}


\begin{center}
{{\bf Stationary Axisymmetric Solutions and their Energy Contents
in Teleparallel Equivalent of Einstein Theory}}\footnote{\hspace{-0.4cm}PACS numbers: 04.20.Cv, 04.20.Fy\\
\hspace*{0.15cm}Keywords:   Teleparallel equivalent of general
relativity,axisymmetric solutions, Gravitational energy-momentum
tensor, extra Hamiltonian.}
\end{center}
\centerline{ Gamal G.L. Nashed\footnote{\hspace{-0.4cm}Mathematics
Department, Faculty of Science, Ain Shams University, Cairo,
Egypt}}

\bigskip

\centerline{\it Centre for Theoretical Physics, The British
University in Egypt,
 El-Sherouk City,}
\centerline{{\it Misr - Ismalia Desert Road, Postal No. 11837,
P.O. Box 43, Egypt.} }

\hspace{2cm}
\\
\\
\\
\\
\\
\\
\\
\\

We apply  the energy-momentum tensor which is coordinate
independent to calculate the  energy content of the axisymmetric
solutions.  Our results are compared with what have been obtained
before within the framework of Einstein general relativity and M\o
ller's tetrad theory of gravitation.

\begin{center}
\newsection{\bf Introduction}
\end{center}

General geometric arena of PGT, the Riemann-Cartan space $U_4$,
may be a priori restricted by imposing certain conditions on the
curvature and the torsion. Thus, Einstein's GR is defined in
Riemann space $V_4$, which is obtained from $U_4$ by the
requirement of vanishing torsion. Another interesting limit of PGT
is the {\it teleparallel or Weitzenb$\ddot{o}$ck} geometry $T_4$.
The vanishing of the curvature means that parallel transport is
path independent. The teleparallel geometry is, in sense,
complementary to Riemannian: curvature vanishes, and torsion
remains to characterize the parallel transport. Of particular
importance for the physical interpretation of the teleparallel
geometry the fact that there is a one-parameter family of
teleparallel Lagrangians which is empirically equivalent to GR
\cite{HNV,HS9,HMM,Nj}. For the parameter value $B=1/2$ the
Lagrangian of the theory coincides, modulo a four-divergence, with
the Einstein-Hilbert Lagrangian, and defines (TEGR).

 The teleparallel
equivalent of general relativity (TEGR) is a viable alternative
geometrical description of Einstein's general relativity written
in terms of the tetrad field \cite{Hf} and continue to be object
of thorough investigations \cite{OP,Yic,Yig,OP6}. In the framework
of the TEGR it has been possible to address the longstanding
problem of defining the energy, momentum and angular momentum of
the gravitational field \cite{MDTC,MFC,MjA}. The tetrad field
seems to be a suitable field quantity to address this problem,
because it yields the gravitational field and at the same time
establishes a class of reference frames in space-time \cite{MVR}.
Moreover there are simple and clear indications that the
gravitational energy-momentum defined in the context of the TEGR
provides a unified picture of the concept of mass-energy in
special and general relativity.

The tetrad formulation of gravitation was considered by M\o ller
in connection with attempts to define the energy of gravitational
field \cite{Mo8,Mo2}. For a satisfactory description of the total
energy of an isolated system it is necessary that the
energy-density of the gravitational field is given in terms of
first- and/or second-order derivatives of the gravitational field
variables. It is well-known that there exists no covariant,
nontrivial expression constructed out of the metric tensor.
However, covariant expressions that contain a quadratic form of
first-order derivatives of the tetrad field are feasible. Thus it
is legitimate to conjecture that the difficulties regarding the
problem of defining the gravitational energy-momentum are related
to the geometrical description of the gravitational field rather
than are an intrinsic drawback of the theory \cite{MDTC,Mj}.

Definition of the angular momentum of the gravitational field is
given in the framework of the TEGR \cite{MUFR}. In similarity to
the definition of the gravitational energy-momentum, Maluf et al.
\cite{MUFR} have interpreted the appropriate constraint equations
as equations that defined the gravitational angular momentum. This
definition turned out to be coordinate independent. The definition
of $P^a$ is invariant under global $SO(3,1)$ transformations. It
has been argued elsewhere \cite{Mj5} that it makes sense to have a
dependence of $P^a$ on the frame. The energy-momentum in classical
theories of particles and fields does depend on the frame, and it
has been asserted that such dependence is a natural property of
the gravitational energy-momentum. The total energy of a
relativistic body, depends on the frame.

A well posed and mathematically consistence expression for the
gravitational energy has been developed \cite{MDTC}. It arises in
the realm of the Hamiltonian formulation of the TEGR \cite{MR} and
satisfies several crucial requirements for any acceptable
definition of gravitational energy. The gravitational
energy-momentum $P^a$ \cite{MDTC,MFC} obtained in the framework of
the TEGR has been investigated in the context of several distinct
configuration of the gravitational filed. For asymptotically flat
space-times $P^{(0)}$ yields the ADM energy \cite{ADM}. In the
context of tetrad theories of gravity, asymptotically flat
space-times may be characterized by the asymptotic boundary
condition \[ e_{a \mu} \cong \eta_{a \mu} + \displaystyle{1 \over
2} h_{a \mu}(1/r),\] and by the condition $\partial_\mu
{e^a}_\mu=O(1/r^2)$ in the asymptotic limit $r \rightarrow
\infty$, with\\ $\eta_{a b}=(-1,+1,+1,+1)$ is the metric of
Minkowski space-time. An important property of tetrad fields that
satisfy the above as that in the flat space-time limit one has
${e^a}_\mu(t,x,y,z)={\delta^a}_\mu$, and therefore the torsion
tensor ${T^a}_{\mu \nu}=0$. Maluf \cite{MVR} has extended the
definition of $P^a$ for the gravitational energy-momentum
\cite{MDTC,MR} to any arbitrary tetrad fields, i.e., for the
tetrad fields that satisfy ${T^a}_{\mu \nu} \neq 0$ for the flat
space-time. The redefinition is the only possible consistent
extension of $P^a$, valid for the tetrad fields that do not
satisfy the above equation.

Recently, Sharif and Amir (2007) have found the teleparallel (TP)
version of the non-null Einstein Maxwell solutions \cite{SJ}.
Then, they have used the TP version of M\o ller (1978) to evaluate
the energy-momentum distribution  of these solutions. They have
found that the energy content in the TP theory is equal to the
energy in GR (as found by Sharif and Fatima (2006)) plus some
additional part. Also they have discussed three possibilities for
the axial-vector field.

It is the aim of the present work to calculate the energy content
of the  axisymmetric solutions using the definition of the
gravitational energy which is coordinate independent. In Sect. 2,
a brief review of the derivation of the field equations of the
gravitational field is given. A summary of the derivation of
energy and angular momentum in TEGR is also given in Sect. 2. In
Sect. 3, we derive the axially symmetric solutions in TEGR and
then, calculate their energy content. The final section is devoted
to discussion and conclusion.
\newsection{The  TEGR for gravitation}

In a spacetime with absolute parallelism the parallel vector field
${e_a}^\mu$ define the nonsymmetric affine connection \be
{\Gamma^\lambda}_{\mu \nu} \stackrel{\rm def.}{=} {e_a}^\lambda
{e^a}_{\mu, \nu}, \ee where $e_{a \mu, \nu}=\partial_\nu e_{a
\mu}$\footnote{spacetime indices $\mu, \ \ \nu, \cdots$ and
SO(3,1) indices a, b $\cdots$ run from 0 to 3. Time and space
indices are indicated to $\mu=0, i$, and $a=(0), (i)$.}. The
curvature tensor defined by ${\Gamma^\lambda}_{\mu \nu}$ is
identically vanishing, however. The metric tensor $g_{\mu \nu}$
 is given by
 \be g_{\mu \nu}= \eta_{a b} {e^a}_\mu {e^b}_\nu, \ee with the
Minkowski metric $\eta_{ a b}=\textrm {diag}(+1\; ,-1\; ,-1\;
,-1)$ \footnote{ Latin indices are rasing and lowering with the
aid of $\eta_{ a b}$ and $\eta^{ a b}$.}.

  The Lagrangian density for the gravitational field in the TEGR,
  in the presence of matter fields, is given by\footnote{Throughout this paper we use the
relativistic units$\;$ , $c=G=1$ and $\kappa={8\pi}$.}
\cite{MDTC,MVR} \be  {\cal L}_G  =  e L_G =- \displaystyle {e
\over 16\pi}  \left( \displaystyle {T^{abc}T_{abc} \over
4}+\displaystyle {T^{abc}T_{bac} \over 2}-T^aT_a
  \right)-L_m= - \displaystyle {e \over 16\pi} {\Sigma}^{abc}T_{abc}-L_m,\ee
where $e=det({e^a}_\mu)$. The tensor ${\Sigma}^{abc}$ is defined
by \be {\Sigma}^{abc} \stackrel {\rm def.}{=} \displaystyle{1
\over 4}\left(T^{abc}+T^{bac}-T^{cab}\right)+\displaystyle{1 \over
2}\left(\eta^{ac}T^b-\eta^{ab}T^c\right).\ee $T^{abc}$ and $T^a$
are the torsion tensor and the basic vector field  defined by \be
{T^a}_{\mu \nu} \stackrel {\rm def.}{=}
{e^a}_\lambda{T^\lambda}_{\mu
\nu}=\partial_\mu{e^a}_\nu-\partial_\nu{e^a}_\mu,\ee and \be
 T^\mu \stackrel {\rm def.}{=} {{T^\nu}_\nu}^\mu, \qquad T^a \stackrel {\rm
def.}{=} {e^a}_\mu T^\mu={{T^b}_b}^a.\ee  The quadratic
combination $\Sigma^{abc}T_{abc}$ is proportional to the scalar
curvature $R(e)$, except for a total divergence term \cite{Mj}.
$L_m$ represents the Lagrangian density for matter fields.

The gravitational field equations for the system described by
${\it L_G}$ are the following
 \be e_{a \lambda}e_{b \mu}\partial_\nu\left(e{\Sigma}^{b \lambda \nu}\right)-e\left(
 {{\Sigma}^{b \nu}}_a T_{b \nu \mu}-\displaystyle{1 \over 4}e_{a \mu}
 T_{bcd}{\Sigma}^{bcd}\right)= \displaystyle{1 \over 2}{\kappa} eT_{a
 \mu},\ee

where \[ \displaystyle{ \delta L_m \over \delta e^{a \mu}} \equiv
e T_{a \mu}.\] It  is possible to prove by explicit calculations
that the left hand side of the symmetric part of the field
equations (7) is exactly given by \cite{MDTC}
 \[\displaystyle{e \over 2} \left[R_{a
\mu}(e)-\displaystyle{1 \over 2}e_{a \mu}R(e) \right]. \] The
axial-vector part of the torsion tensor $A_\mu$ is defined by \be
A_\mu \stackrel{\rm def.}{=} {1 \over 6} \epsilon_{\mu \nu \rho
\sigma} T^{\nu \rho \sigma}={1 \over 3} \epsilon_{\mu \nu \rho
\sigma} \gamma^{\nu \rho \sigma}, \qquad where \qquad
\epsilon_{\mu \nu \rho \sigma} \stackrel{\rm def.}{=} \sqrt{-g}
\delta_{\mu \nu \rho \sigma}, \ee with $\gamma_{\nu \rho
\sigma}=\eta^{a b}e_{a \nu} e_{b \rho\; ;\; \sigma}$ being the
contorsion tensor and $\delta_{\mu \nu \rho \sigma}$ is completely
antisymmetric and normalized as $\delta_{0123}=-1$.

In the context of Einstein's general relativity, rotational
phenomena is certainly not a completely understood issue. The
prominent manifestation of a purely relativistic rotation effect
is the dragging of inertial frames. If the angular momentum of the
gravitational field of isolated system has a  meaningful notion,
then it is reasonable to expect the latter to be somehow related
to the rotational motion of the physical sources.

The angular momentum of the gravitational field has been addressed
in the literature by means of different approaches. The oldest
approach is based on pseudotensors \cite{BT,LL}, out of which
angular momentum superpotentials are constructed. An alternative
approach assumes the existence of certain Killing vector fields
that allow the construction of conserved integral quantities
\cite{Ka}. Finally, the gravitational angular momentum can also be
considered in the context of Poincar$\acute{e}$ gauge theories of
gravity \cite{HS8}, either in the Lagrangian or in the Hamiltonian
formulation. In the latter case it is required that the generators
of spatial rotations at infinity have a well defined functional
derivatives. From this requirement a certain surface integral
arises, whose value is interpreted as the gravitational angular
momentum.

The Hamiltonian formulation of TEGR is obtained by establishing
the phase space variables. The Lagrangian density does not contain
the time derivative of the tetrad component $e_{a0}$. Therefore,
this quantity will arise as a Lagrange multiplier \cite{Dp}. The
momentum canonically conjugated to $e_{ai}$ is given by
$\Pi^{ai}=\delta L/\delta \dot{e}_{ai}$. The Hamiltonian
formulation is obtained by rewriting the Lagrangian density in the
form $L=p\dot{q}-H$, in terms of $e_{ai}, \Pi^{ai}$ and the
Lagrange  multipliers. The Legendre transformation can be
successfully carried out and the final form of the Hamiltonian
density has the form \cite{MR} \be H=e_{a0}C^a+\alpha_{ik}
\Gamma^{ik}+\beta_k\Gamma^k,\ee plus a surface term. Here
$\alpha_{ik}$ and $\beta_k$ are Lagrange multipliers that are
identified as \be \alpha_{ik}={1 \over 2} (T_{i0k}-T_{k0i}) \qquad
and \qquad \beta_k=T_{00k},\ee and $C^a$, $\Gamma^{ik}$ and
$\Gamma^k$ are first class constraints. The Poisson brackets
between any two field quantities $F$ and $G$ is given by \be \{
F,G \}=\int d^3x \left( \displaystyle{\delta F \over \delta
e_{ai}(x)} \displaystyle{\delta G \over \delta
\Pi^{ai}(x)}-\displaystyle{\delta F \over \delta
\Pi^{ai}(x)}\displaystyle{\delta G \over \delta e_{ai}(x)}
\right).\ee We recall that the Poisson brackets
$\left\{\Gamma^{ij}(x),\Gamma^{kl}(x)\right\}$ reproduce the
angular momentum algebra \cite{Mj}.

 The constraint $C^a$ is
written as $C^a=-\partial_i \Pi^{ai}+h^a$, where $h^a$ is an
intricate expression of the field variables. The integral form of
the constraint equation $C^a=0$ motivates the definition of the
gravitational energy-momentum $P^a$ four-vector \cite{Mj} \be
P^a=-\int_V d^3 x
\partial_i \Pi^{ai},\ee where $V$ is an arbitrary volume of the
three-dimensional space. In the configuration space we have \be
\Pi^{ai} = -\frac{2}{\kappa} \sqrt{-g} \Sigma^{a0i}, \ee with \[
\partial_\nu(\sqrt{-g}\Sigma^{a \lambda \nu})=\displaystyle{\kappa
\over 2}\sqrt{-g}{e^a}_\mu (t^{\lambda \mu}+T^{\lambda \mu}),
\quad where \quad   t^{\lambda \mu}=\frac{1}{2\kappa}
\left(4\Sigma^{bc \lambda}{T_{bc}}^\mu-g^{\lambda \mu}
\Sigma^{bcd}T_{bcd} \right).\]

The emergence of total divergences in the form of scalar or vector
densities is possible in the framework of theories constructed out
of the torsion tensor. Metric theories of gravity do not share
this feature. By making $\lambda=0$ in Eq. (13) and identifying
$\Pi^{ai}$ in the left side of the latter, the integral form of
Eq. (13) is written as \be P^a=\int_V d^3 x \sqrt{-g}
{e^a}_\mu\left(t^{0 \mu}+T^{0 \mu} \right).\ee Eq. (14) suggests
that $P^a$ is now understood as the gravitational  energy-momentum
\cite{Mj5}. The spatial component $P^{(i)}$ form a total
three-momentum, while temporal component $P^{(0)}$ is the total
energy \cite{LL}.

It is possible to rewrite the Hamiltonian density of Eq. (9) in
the equivalent form \cite{MUFR} \be H=e_{a0}C^a+\displaystyle{1
\over 2}\lambda_{ab}\Gamma^{ab}, \qquad with \qquad
\lambda_{ab}=-\lambda_{ba}, \ee are the Lagrangian multipliers
that are identified as $\lambda_{ik}=\alpha_{ik}$ and
$\lambda_{0k}=-\lambda_{k0}=\beta_k$.  The constraints
$\Gamma^{ab} = -\Gamma^{ba}$ \cite{MR} embodies both constraints
$\Gamma^{ik}$ and $\Gamma^k$ by means of the relation \be
\Gamma^{ik}={e_a}^i {e_b}^k \Gamma^{ab}, \qquad and \qquad
\Gamma^k \equiv \Gamma^{0k}={e_a}^0 {e_b}^k \Gamma^{ab}.\ee The
constraint $\Gamma^{ab}$ can be reads as \be
\Gamma^{ab}=M^{ab}+\frac{2}{\kappa}\sqrt{-g}{e_{c}}^0
\left(\Sigma^{acb}-\Sigma^{bca}\right).\ee

In similarity to the definition of $P^a$, the integral form of the
constraint equation $\Gamma^{ab}=0$ motivates the new definition
of the space-time angular momentum. The equation $\Gamma^{ab}=0$
implies \be M^{ab}=-\frac{2}{\kappa}\sqrt{-g}{e_{c}}^0
\left(\Sigma^{acb}-\Sigma^{bca}\right).\ee Maluf et al. \cite{Mj,
MUFR} defined \be L^{ab} = 2\int_V d^3x M^{[ab]}, \ee  as the
4-angular-momentum of the gravitational field for an arbitrary
volume V of the three-dimensional space. In Einstein-Cartan type
theories there also appear constraints that satisfy the Poisson
bracket given by Eq. (11). However, such constraints arise in the
form $\Pi^{[ij]}=0$, and so a definition similar to Eq. (19),
i.e., interpreting the constraint equation as an equation for the
angular momentum of the field, {\it is not possible}. Definition
(19) is three-dimensional integral. The quantities $P^a$ and
$L^{ab}$ are separately invariant under general coordinate
transformations of the three-dimensional space and under time
reparametrizations, which is an expected feature since these
definitions arise in the Hamiltonian formulation of the theory.
Moreover, these quantities transform covariantly under global
$SO(3,1)$ transformations \cite{MUFR}.

\newsection{Energy  content of axisymmetric solutions}

Now we are going to calculate the energy content of the
axisymmetric  tetrad field that has the form \cite{SJ} \be
\left({e_i}^{  \mu} \right) = \left( \matrix{ 1& 0 & 0 & 0
 \vspace{3mm} \cr  \displaystyle{B(\rho\; ,z) \over F(\rho)}
 \sin\phi & e^{-K(\rho,z)} \cos\phi &-\displaystyle{1 \over F(\rho)}
 \sin\phi & 0 \vspace{3mm} \cr -\displaystyle{B(\rho\; ,z) \over F(\rho)}
 \cos\phi & e^{-K(\rho,z)} \sin\phi &\displaystyle{1 \over F(\rho)}
 \cos\phi & 0   \vspace{3mm} \cr 0&0 & 0 &e^{-K(\rho,z)}    \cr }
\right)\; . \ee Using Eq. (2) the associated metric of the tetrad
field given by Eq. (20) takes the well known form \be
ds^2=dt^2-e^{2K(\rho,z)}d\rho^2-\left(F^2(\rho)-B^2(\rho\;
,z)\right)d\phi^2-e^{2K(\rho,z)} dz^2+2B(\rho\; ,z)dtd\phi,\ee
 $B(\rho\; ,z)$, $K(\rho,z)$ and  $F(\rho)$ are unknown
functions which satisfy the following relations \ba \dot{B} \A=\A
FW', \qquad \qquad B'=-\displaystyle{1 \over 4}
aF\left(\dot{W}^2-W'^2\right), \nonu
K' \A=\A -\displaystyle{1 \over 2} aF\dot{W}W', \qquad \qquad
\ddot{W}+\dot{F}F^{-1}\dot{W}+W''=0, \ea where dot and prime
denoting the derivatives w.r.t. $\rho$ and $z$ respectively. Here
$a$ is a constant and $W$ is an arbitrary function of $\rho$ and
$z$ in general \cite{SJ}.  McIntosh's give a solution in the form
$W=-2bz$ while  McLenaghan et. al. solution's has the form
$W=2ln\rho$ \cite{TB}. The above metric represents a five classes
of non-null electromagnetic field and prefect fluid solutions
which possesses a metric symmetry not inherited by the
electromagnetic field and admits a homothetic vector field. Two of
these classes  contain electrovac solutions as special cases,
while the other three necessarily contain some fluid.
Generalization of metric given by Eq. (21) is given in
\cite{SKMH}.

Applying the tetrad field of Eq. (20) to the field equations (7)
we get the non-vanishing components to have the form \ba {T^0}_0
\A =\A \displaystyle{-e^{-2K} \over 64\pi
F^3}\left(2B[F\ddot{B}+FB''-\dot{B}\dot{F}]+3F[\dot{B}^2+B'^2]-4F^2[F\ddot{K}+FK''+\ddot{F}]\right),
\nonu
{T^0}_2 \A =\A \displaystyle{-e^{-2K} \over 32\pi
F^3}\left(F^2[F\ddot{B}+FB''-\dot{B}\dot{F}-2B\ddot{F}]-B^2[\dot{B}\dot{F}-F\ddot{B}-FB'']+2BF[
\dot{B}^2+B'^2]\right),\nonu
{T^1}_1 \A =\A \displaystyle{e^{-2K} \over 64\pi
F^2}\left(\dot{B}^2+4F\dot{F}\dot{K}-B'^2\right),\quad \qquad
{T^1}_3={T^3}_1 = \displaystyle{e^{-2K} \over 32\pi
F^2}\left(\dot{B}B'+2F\dot{F}K'\right),\nonu
 {T^2}_0 \A =\A
\displaystyle{e^{-2K} \over 32\pi
F^3}\left(F[\ddot{B}+B'']-\dot{B}\dot{F}\right),\quad  {T^2}_2 =
\displaystyle{e^{-2K} \over 64\pi
F^3}\left(F[\dot{B}^2+2B\ddot{B}+B'^2+2BB'']-2B\dot{B}\dot{F}+4F^3[\ddot{K}+K'']\right),\nonu
{T^3}_3 \A=\A \displaystyle{e^{-2K} \over 64\pi
F^2}\left(B'^2-\dot{B}^2+4F[\ddot{F}-\dot{F}\dot{K}]\right),
 \ea where ${T^\mu}_\nu$ is the energy-momentum tensor.

A special solution of the above non linear P.D.E can be obtained
by choosing \cite{SJ}\\  \be  B(\rho,z)=\displaystyle{m \over
n}e^{n\rho}, \qquad F(\rho)=e^{n\rho}, \qquad K(\rho,z)=0,\ee
where $m$ and $n$ are constants. The above solution reproduce the
well known solution which is known as the electromagnetic
generalization of the G$\ddot{o}$del solution \cite{SJ,TB}. \\ The
second solution  can be obtained by choosing \be B(\rho,z)=e^{a
\rho}, \qquad F(\rho)=\displaystyle{e^{a \rho} \over \sqrt{2}},
\qquad K(\rho,z)=0,\ee which known as the space time homogenous
G$\ddot{o}$del metric \cite{SJ,TB}.

Now let us calculate the energy content of the tetrad (20) using
(12). To do so let us calculate the non-vanishing components of
the torsion tensor. Using Eq. (20) in Eq. (5) we get \ba
{T^{(0)}}_{12}\A=\A \displaystyle{1 \over F}\left((
e^{K}-\dot{F})[e^K BF \sin^2\phi
+B(F\cos\phi+1)]-F\dot{B}\right),\quad {T^{(0)}}_{13}= -e^{2K}
BK'\sin \phi \cos\phi  \nonu
{T^{(0)}}_{23}\A=\A B', \quad
{T^{(1)}}_{12}=\sin\phi(\dot{F}-e^K),\quad {T^{(1)}}_{13}=
 e^K K'\cos\phi ,\quad
{T^{(2)}}_{12} = (e^{K} -\dot{F})(e^K \sin^2\phi+\cos\phi),\nonu
 {T^{(2)}}_{13}\A=\A -e^{2K} K'\sin\phi \cos\phi, \quad {T^{(3)}}_{13}=-e^k \dot{K}.\ea
 The non-vanishing components of the basic vector $T_\mu$ are
 \be T_1=(\dot{F}-e^K)(e^K \sin^2 \phi +\cos
 \phi)+e^K\dot{K}, \qquad T_2=\sin\phi(\dot{F}-e^K),
 \qquad T_3=e^K K' \cos \phi. \ee

 Pereira et al. \cite{PZ} have proved that the axial vector tensor plays the
 role of the gravitomagnetic component of the gravitational field
 in the case of slow rotation and weak field approximations. The non-vanishing
 components of the axial vector tensor, defined by Eq. (8), associated with the
  tetrad field given by Eq. (20) are
 \ba  A^0 \A=\A \displaystyle{K'\sin \phi \cos \phi(F^2-2B^2) \over
 3F},\qquad \qquad A^1 = \displaystyle{e^{-2K}B' \over 3F},
 \qquad \qquad A^2=\displaystyle{2BK'\sin\phi \cos \phi \over 3F}, \nonu
 A^3 \A=\A \displaystyle{1 \over
 3F^2}\left(\left[2BF(1-e^{-K}\dot{F})\{\sin^2\phi+e^{-K}\cos\phi\}
 +Be^{-K}\right]-e^{-2K}(F\dot{B}+B\dot{F})\right).\ea
{\it It is of interest to compare our results with that obtained
before by Sharif and Amir (2007). They have calculated the axial
vector of the tetrad given by Eq. (20) and wrote  the
non-vanishing components as  $A^1$ which coincides with what we
have obtained in Eq. (28) and the other component as $A^3$ that
has the form \be A^3=\displaystyle{\dot{B}e^{-2K} \over 3F},\ee
which is completely different from that obtained in Eq. (28).
Therefore, the analysis related to the axial vector part given in
Ref. \cite{SJ} will completely now need some modifications which
we will do. The spacelike axial vector can now be written}
\cite{SJ} \be {\bf A}=\sqrt{-g_{11}}A^1{\hat
e}_\rho+\sqrt{-g_{22}}A^2{\hat e}_\phi+\sqrt{-g_{33}}A^3{\hat
e}_z,\ee where ${\hat e}_\rho$, ${\hat e}_\phi$ and ${\hat e}_z$
are unit vectors along the radial $\rho$, $\phi$ and z-directions
respectively. Using Eq. (28) in (30) we get \ba {\bf A} \A=\A
\displaystyle{e^{-K}B' \over 3F}{\hat
e}_\rho+\displaystyle{2\sqrt{F^2-B^2} BK'\sin\phi \cos \phi \over
3F}{\hat e}_\phi\nonu
\A \A+ \displaystyle{e^K \over
3F^2}\left((1-e^{-K}\dot{F})\left[2BF\{\sin^2\phi+e^{-K}\cos\phi\}
 +Be^{-K}\right]-e^{-2K}F\dot{B}\right){\hat e}_z.\ea
The spin precession of a Dirac particles in teleparallel gravity
is related to the axial vector by \cite{SJ,HS2} \be
\displaystyle{d{\bf S} \over dt}=-{\bf b}\times {\bf S},\ee where
${\bf S}$ is the spin vector of a Dirac particles and ${\bf b}=3/2
{\bf A}$ where ${\bf A}$ is given by Eq. (31).

The direct evidence for the coupling of intrinsic spin to the
rotation of the Earth has become available \cite{Mbl}. According
to the TEGR every spin $\frac{1}{2}$ particle in the laboratory
has an additional interaction Hamiltonian. However, such
 intrinsic spin must precess in a sense opposite to the sense of
 rotation of the Earth as measured by the
 observer. The corresponding
extra Hamiltonian associated with such motion  would be of the
form \cite{Mb} \be \delta H=-{\bf b} \cdot \sigma,\ee where
$\sigma$ is the spin of the particle.

To calculate the energy density associated with the tetrad field
given by Eq. (20) we must calculate $\Sigma^{\mu \nu \lambda}$
which defined in Eq. (4). The necessary non-vanishing components
of $\Sigma^{\mu \nu \lambda}$ are \ba \Sigma^{001}\A=\A
\displaystyle{1 \over 2F^3}\Biggl((1-e^{-K}\dot{F})\{F(\sin^2
\phi+e^{-K}\cos\phi)(B^2-F^2)+e^{-K}B^2\}  -e^{-2K}BF\dot{B}
+e^{-K}F^3\dot{K}\Biggr), \nonu
\Sigma^{002}\A=\A \displaystyle{-\sin\phi(e^K-\dot{F}) \over
2F^2}, \quad  \Sigma^{003}= \displaystyle{-e^{-K}(e^{-K}BB'-\cos
\phi F^2K') \over 2F^2}, \nonu
\Sigma^{101}\A=\A \displaystyle{-B\sin\phi
(e^{-K}-e^K-(e^{-2K}-1)\dot{F}) \over 2F^2},\nonu
\Sigma^{102}\A=\A \displaystyle{1 \over
 4F^3}\left((1-e^{-K}\dot{F})\{2BF[\sin^2\phi+e^{-k}\cos\phi]+
 e^{-K}B\}-e^{-2K}F\dot{B}\right),\nonu
\Sigma^{103}\A=\A \displaystyle{-1 \over
 4}e^{-2K}B K'\sin\phi \cos\phi, \quad
\Sigma^{201}= \displaystyle{-e^{-K} \over
 4F^3}(B-Be^{-K}\dot{F}-e^{-K}F\dot{B}),
 \nonu
  \Sigma^{202}\A=\A \displaystyle{B \sin\phi \over
 2F^2}\left(e^{-K}-\dot{F}) \right),\quad
 \Sigma^{203} = \displaystyle{e^{-2K} B' \over
 4F^2}, \quad \Sigma^{301} = \displaystyle{e^{-2K}B K' \sin \phi \cos \phi \over
 4} \nonu
 \Sigma^{302} \A=\A \displaystyle{-e^{-2K}B' \over
 4F^2}, \quad  \Sigma^{303}= \displaystyle{B\sin\phi \over
 2F^2}(e^K-\dot{F}).
 \ea Using (34) in (13) we get
 \ba \Pi^{(0) 0} \A=\A \displaystyle{1 \over
 \kappa
 F^2}\Biggl(F\left\{\frac{d}{d\rho}(B\dot{B})+\frac{d}{dz}(BB')+
 \sin^2\phi\left[B^2-F^2 \right]
 \frac{d}{d\rho}(\dot{F}e^K)\right\}+B\dot{B}\dot{F}\left[1+2F\cos\phi\right]
 +B^2\left[F\ddot{F}-2\dot{F}^2\right]\nonu
 \A \A-\cos\phi F^3\frac{d}{dz}(K'e^K) +\left[\cos\phi
\frac{d}{d\rho}(F\dot{F})+e^K\dot{F}^2\left\{1+F\cos^2\phi\right\}\right]
 \left\{B^2+F^2\right\}+2e^K\left(\frac{B^3}{F}\right)\frac{d}{d\rho}\left(\frac{F}{B}\right)
 \nonu
 \A \A+F^2\frac{d}{d\rho}(Fe^{2K})-e^K\dot{K}\left[B^2+F^2\dot{F}\right]
 -F^2 \cos \phi\left[\frac{d}{d\rho}\left\{Fe^K\left[1+\cos\phi
 e^{K}\right]\right\}\right]+2 \cos \phi \frac{d}{d\rho}(e^KB^2F)
 \nonu
\A \A-F^3\frac{d}{d\rho}( \dot{K}e^K )+e^{K}
 B\sin^2\phi\left[2F\dot{F}\dot{B}-e^K\left\{FB\dot{K}+2F\dot{B}-\dot{F}\right\}\right]
 +e^{3K}F\cos\phi\left[1-e^{-K} \dot{F}\right]
 \Biggr).  \ea
 The non-vanishing components of the momentum density have the
 form
 \ba \Pi^{(1) 0} \A=\A \frac{\sin \phi \cos \phi e^K}{\kappa
 F^3}\Biggl(F\frac{d}{d\rho}\left(Be^K\right)-Be^K\dot{F}-F\frac{d}{d\rho}
 (e^{3K}B)+Be^{3K}\dot{F}+B\dot{F}^2+BF\frac{d}{d\rho}\left(e^{2K}\dot{F}\right)-Be^{2K}\dot{F}^2
 \nonu
 \A \A-F\frac{d}{d\rho}(B\dot{F})+Fe^{2K}\dot{B}\dot{F} -2e^{2K}FB\cos\phi
+2FBe^K \cos \phi \dot{F}+FBe^K-FB\dot{F} +\frac{F^3}{2}\cos\phi
\frac{d}{dz}(BK') \Biggr)\nonu
  \Pi^{(2) 0} \A=\A \frac{\cos \phi}{\kappa
 F^2}\Biggl(Be^K\dot{F}-B\dot{F}^2-\frac{F}{2}(Be^K\dot{K}-e^K\dot{B}+B\ddot{F}+F\ddot{B}+FB'')
 +F^2Be^{3K}\cos\phi-F^2Be^{2K} \cos \phi \dot{F}\Biggr)\nonu
 \Pi^{(3) 0} \A=\A \frac{2\sin \phi \ e^K}{\kappa
 F}\Biggl(\dot{F}e^{2K}[B'+3BK']-\frac{F \cos \phi}{2} (FB
 \dot{K}K'+B\dot{F}K'+F\dot{B}K'+FB\dot{K}')\nonu
\A \A - B e^{3K}(B'+4K')\Biggr).\ea It is of interest to note that
if $(F(\rho)=const, \ K(\rho,z)=K_1(\rho), \ B(\rho,z)=B(\rho))$
then the component of the  momentum density $\Pi^{(3)0}=0$.

Now let us  repeat the same calculations using the solution given
by Eq.(24), in this case the basic vector has the non-vanishing
components \be T_1=(ne^{n\rho}-1)( \sin^2 \phi +\cos \phi), \qquad
T_2=\sin\phi(ne^{n\rho}-1),\ee and the non-vanishing components of
the axial
 vector field are
 \be A^3 =\displaystyle{1 \over
 3}\left(\left[2\frac{m}{n}(1-ne^{n\rho})\{\sin^2\phi+\cos\phi\}
 +\frac{m}{n}e^{-n\rho}\right]-2m\right).\ee
 \be {\bf A} = \displaystyle{1 \over
 3}\left(\left[2\frac{m}{n}(1-ne^{n\rho})\{\sin^2\phi+\cos\phi\}
 +\frac{m}{n}e^{-n\rho}\right]-2m\right) {\hat e}_z.\ee
 The corresponding extra Hamiltonian \cite{ZB} is given by
  \be \delta H =-{\bf b} \cdot
\sigma=-\displaystyle{1 \over
 2}\left(\left[2\frac{m}{n}(1-ne^{n\rho})\{\sin^2\phi+\cos\phi\}
 +\frac{m}{n}e^{-n\rho}\right]-2m\right) {\hat e}_z \cdot
\sigma, \ee
 and the component of the energy density is given by
 \ba \Pi^{(0) 0} \A=\A \displaystyle{e^{n\rho} \over
 n \kappa}\Biggl((n^2-m^2)[\sin^2\phi
 +\cos\phi](1-2ne^{n\rho})+
 ne^{-n\rho}\cos\phi\{e^{-n\rho}-n\}+2m^2n\Biggr). \ea
  The non-vanishing components of the
momentum density have the form \ba
 \Pi^{(1) 0} \A=\A \frac{m\sin \phi e^{n\rho} }{n
\kappa}\left(e^{-n\rho}\{1-3\cos^2\phi-n\}+\frac{e^{-2n\rho}}{2}(1-n)+n(3\cos^2\phi-1)\right)
\nonu
\Pi^{(2) 0} \A=\A \frac{m e^{n\rho}}{\kappa
n^2}\Biggl(m^2\left\{4ne^{n\rho}+3ne^{2n\rho}-2e^{n\rho}\cos\phi-2e^{n\rho}+2e^{n\rho}\cos^2\phi+3n
\cos\phi e^{2n\rho}(1 -\cos\phi)-1\right\}\nonu
 \A \A +n^2\Biggl\{1-3ne^{2n\rho}+n\sin^2\phi+2e^{n\rho}-2\cos^2\phi-\cos\phi+e^{n\rho}\cos\phi
 +3n \cos\phi e^{2n\rho}(\cos\phi-1)-n\cos\phi \nonu
\A \A +e^{n\rho}\cos^2\phi(4\cos^2 \phi
 -5)\Biggr\}+ n\left\{3\cos^2\phi-4\cos^4\phi+e^{-n\rho}(\cos^2\phi-\frac{1}{2})+\cos
\phi(1+e^{-n\rho})\right\} \Biggr) \nonu
 \Pi^{(3) 0} \A=\A 0. \ea
For the  solution given by Eq. (25) the basic vector has the
non-vanishing components \be
T_1=(\frac{1}{\sqrt{2}}ae^{a\rho}-\sqrt{2})( \sin^2 \phi +\cos
\phi), \qquad
T_2=\frac{\sin\phi}{\sqrt{2}}(ae^{a\rho}-\sqrt{2}),\ee and the
non-vanishing components of the axial
 vector field are
 \be A^3 = \displaystyle{1 \over
 3}\Biggl(\left[2 \sqrt{2}e^{-a\rho}+e^{2\rho}(e^{2\rho}+1)(\sqrt{2}-a)\right]\{\sin^2\phi
 +\cos\phi\}-2e^{\rho(4-a)}(\sqrt{2}a
 -e^{-a \rho}) \Biggr).\ee
 \be {\bf A} = \displaystyle{1 \over
 3}\Biggl(\left[2 \sqrt{2}e^{-a\rho}+e^{2\rho}(e^{2\rho}+1)(\sqrt{2}-a)\right]\{\sin^2\phi
 +\cos\phi\}-2e^{\rho(4-a)}(\sqrt{2}a
 -e^{-a \rho}) \Biggr){\hat e}_z.\ee
 The corresponding extra Hamiltonian \cite{ZB} is given by \be \delta H =-{\bf b} \cdot
\sigma=-\displaystyle{1 \over
 2}\Biggl(\left[2 \sqrt{2}e^{-a\rho}+e^{2\rho}(e^{2\rho}+1)(\sqrt{2}-a)\right]\{\sin^2\phi
 +\cos\phi\}-2e^{\rho(4-a)}(\sqrt{2}a
 -e^{-a \rho})\Biggr) {\hat e}_z \cdot \sigma, \ee and the
 component of the energy density has the form
 \be \Pi^{(0) 0}= \displaystyle{e^{a\rho} \over
  \kappa}\Biggl( \sin^2\phi
 (a^2 e^{a\rho}-\frac{a}{\sqrt{2}})+\cos\phi(a^2e^{a\rho}-\frac{a}{\sqrt{2}})
 +\frac{a^2}{\sqrt{2}}+e^{-a\rho} \cos\phi (\sqrt{2}e^{-a\rho}-a)
 \Biggr).\ee
The non-vanishing components of the momentum density have the form
\ba \Pi^{(1) 0} \A=\A \frac{\sin \phi e^{a\rho} }{
\kappa}\left(e^{-2a\rho}
\{1-a\}+e^{-a\rho}(1-a\sqrt{2}-3\sqrt{2}\cos^2\phi)+a(3\cos^2\phi-1)
\right) \nonu
\Pi^{(2) 0} \A=\A \frac{ e^{a\rho}}{2\kappa
}\Biggl(e^{-a\rho}(2\sqrt{2}\cos\phi(1+\cos\phi)-\sqrt{2})+\sqrt{2}e^{a\rho}(8a^2-a\cos^2\phi
-3a\cos\phi+4a\cos^4\phi-2a) \nonu
 \A \A +3a^2e^{2a\rho}(\sin^2\phi+\cos\phi)+2\cos^2\phi(3-4\cos^2\phi-a^2-2a)+
 2\cos\phi(1-a-a^2)+2a(a-1) \Biggr), \nonu
 \Pi^{(3) 0} \A=\A 0. \ea

\newsection{Main results and discussion}

We have applied the axisymmetric tetrad field given by Eq. (20)
with three unknown functions of $\rho$ and $z$  to the field
equations (7).  We have obtained two special solutions by taking
particular values of the unknown functions $E$, $F$ and $K$. The
first one is given by Eq. (24) and is known as {\it
electromagnetic generalization of the G$\ddot{o}$del solution}
\cite{Tj}, while the second one is given by Eq. (25)  and is known
as {\it G$\ddot{o}$del solution} \cite{Tj}. These solutions are
special solutions of the non-linear P.D.E. given by Eq. (23)
\cite{SJ,Tj}.

We have calculated the basic vector $T^a$ defined by Eq. (6) for
the tetrad field (20). The components  we have obtained  as given
by Eq. (27) are different from the components obtained in Ref.
\cite{SJ} which are \be T_1=\frac{-1}{F}(\dot{F}-e^K)-\dot{K},
\qquad \qquad T_3=-K'.\ee We have calculated the axial vector part
applied Eq. (20) to Eq. (8). Our results are completely different
from the results obtained in Ref. \cite{SJ}.  {\it Therefore, the
analysis of the extra Hamilton used in \cite{SJ} is again
discussed. The extra Hamiltonian is now recalculated for the
tetrad field of Eq. (20) and is given by Eq. (33).}

The energy  density and   momentum density are calculated using
the energy-momentum  tensor of TEGR which is coordinate
independent. For the energy
 momentum density and momentum density derived in Ref. \cite{SJ}
 we have some comments:\\ i) The energy
 momentum density and momentum density, are not correct.\vspace{.2cm}\\ ii) In Ref. \cite{SJ} a discussion is given
 for the case when the arbitrary parameter
 $\lambda=1$.  It is not clear whey $\lambda=1$ is discussed in
 spite that for this case still the theory of M\o ller  deviates
 from GR. \vspace{.2cm}\\ iii) When the tetrad (21) of Ref. \cite{SJ} applied
 to the superpotential (12) of \cite{SJ}, it is logic to obtain
 the energy density and momentum density with the arbitrary parameter $\lambda$. If we
 want to compare the result with GR we must take $\lambda=0$ not $\lambda=1$!
 {\it  Here in this work we have done  the
 calculations  of the energy density and momentum density using the
  gravitational energy-momentum tensor which is coordinate independent
  since it is constructed out from Hamilton structure}.

\end{document}